\documentclass{aastex631}
\usepackage{multirow}
\usepackage{gensymb}
\usepackage{xcolor}
\newcommand{\mmhb}{$M_{\rm mmhb}~$}
\newcommand{\mmax}{$M_{\rm max}~$}
\newcommand{\tc}{}
\newcommand{\te}{}
\begin{document}
\title{Stable Hydrogen burning limits in rapidly rotating \tc{very low mass objects}} 
\correspondingauthor{Tapobrata Sarkar}
\email{tapo@iitk.ac.in}

\author{Shaswata Chowdhury}
\affiliation{Department of Physics, Indian Institute of Technology,
	Kanpur 208016, India}

\author{Pritam Banerjee}
\affiliation{Department of Physics, Indian Institute of Technology,
	Kanpur 208016, India}

\author{Debojyoti Garain}
\affiliation{Department of Physics, Indian Institute of Technology,
	Kanpur 208016, India}

\author{Tapobrata Sarkar}
\affiliation{Department of Physics, Indian Institute of Technology,
	Kanpur 208016, India}

\begin{abstract}
We present novel effects of uniform rapid stellar rotation on the minimum mass of stable 
hydrogen burning in very low mass stars, using an analytic model, and relaxing the 
assumption of spherical symmetry. We obtain an analytic formula for the 
minimum mass of hydrogen burning as a function of the \tc{angular speed of stellar rotation}. Further, we show the existence of 
a maximum mass of stable hydrogen burning in such stars, which is purely an artefact of rapid rotation. 
The existence of this extremum in mass results in a \tc{minimum admissible value of the stellar rotation period $\sim 22$ min},
\tc{below} which \tc{a very low mass object does not reach the main sequence}, within the ambits of our model. For a given angular speed, we
predict a mass range beyond which \tc{such an object will not evolve into a main sequence star.} 
\end{abstract}

\section{Introduction}

Brown dwarfs (BDs), which were theoretically predicted by \cite{SSK}, and \cite{Hayashi}, are substellar objects whose masses
range between \tc{thirteen times that of Jupiter ($\sim 10^{-2}M_{\odot}$) }
and stars at the bottom of the main sequence ($\sim 10^{-1} M_{\odot}$). During their lifetimes, these ``failed stars'' do 
not attain sustained nuclear fusion of Hydrogen into Helium, as their masses are less than a certain 
minimum value, dubbed as the minimum mass of hydrogen burning ($M_{\rm mmhb}$) or sometimes as the minimum main sequence mass. 
\cite{BL} review the works in the area from the mid 60's to the early 90's (see also \cite{Evol1}, \cite{Evol2}), 
and provide analytical models of BDs and very low mass \tc{(VLM)} stars, 
although their observational aspects were still in the nascent stages at that time, given that BDs are particularly 
difficult to detect, due to their typically low luminosities. Later, \cite{Rebolo} announced the first observation of 
a brown dwarf in the Pleiades cluster, and this was closely followed by the similar discovery of \cite{Nakajima}. 
The plethora of activities that followed immediately thereafter, are well documented in the review articles by 
\cite{CB}, \cite{Basri}, \cite{BL2}, 
and the textbook by \cite{Rebolo2} (see also \cite{Evol3}, \cite{Burrows2}, \cite{CB2}). 
The research carried out in the area in the next decade is outlined in the more recent textbook by \cite{Joergens}
(see also \cite{Allard}, \cite{Chabrier}, \cite{Marley}).

What distinguishes BDs from \tc{VLM main sequence stars} is the $M_{\rm mmhb}$, with the currently accepted value of $\sim 0.08 M_{\odot}$,
assuming a static scenario (the recent review of \cite{Auddy} quotes the range $0.064 - 0.087 M_{\odot}$, 
based on some modifications of earlier analytical models). 
However, it is by now well known that various factors may affect the $M_{\rm mmhb}$, one example being accretion 
in binary systems (\cite{SalpeterAccretion}). In this context, we show here that the $M_{\rm mmhb}$ can also be enhanced from its accepted
value, via stellar rotation (the physics of rotating stars are described in sufficient details in the older literature,
e.g. \cite{KThomas} and in the recent monographs by 
\cite{Tassoul} and \cite{Maeder}). Indeed, more than five decades ago, \cite{KippenRot} showed a possible 
increase in the $M_{\rm mmhb}$ due to rotational effects. The basic physics may seem simple. Namely, that with centrifugal
forces effectively reducing gravity inside a stellar object, a rotating star can \tc{maintain hydrostatic equilibrium at lower core densities and temperatures, thus requiring more mass to achieve thermal stability than its non-rotating
cousin.} Here one has to keep in mind that in rotating stellar objects, all stellar parameters like the density,
temperature etc., depend on the \tc{angular speed of rotation} $\Omega$, and that there are several competing effects involving the degeneracy
as well. One of the results in this paper is an analytic formula of $M_{\rm mmhb}$ as a function of $\Omega$.

In particular, we consider rapid rotation, where the approximation of spherical symmetry needs to be abandoned. 
By rapid, we mean a rotation period much smaller than that of Jupiter, which has a period of $\sim 10~{\rm h}$. 
Such rapid rotations in cool dwarfs have been abundantly reported in the recent past. \cite{Clarke} presented photometric observations
of a T6 dwarf with a rotation period of $1.41~{\rm h}$. \cite{Metchev} presented data on a T7 dwarf with
a rotation period of $1.55~{\rm h}$. The analysis of \cite{Route} obtains a dramatically smaller 
period of $\sim 17$ min for a T6 dwarf, although the authors point out that this might be a subharmonic of a longer period. 
Followup observations of the same object by \cite{Williams} however indicated that this period might in fact be closer
to $1.93$ h, although these authors also mention the need for more data to confirm this. The most recent analysis appears 
in \cite{Tannock}, who reported on the observation of photometric periods ranging from $1.08$ to $1.23$ h. Clearly then, the latest
available data on the rapidly rotating brown dwarfs point to the smallest period of $1.08$ h, and \cite{Tannock}
claim that these are ``unlikely'' to rotate much faster, given the clustering of the BDs having the smallest rotation periods.  

With this status of observational signatures, the question we ask here is, if there are any constraints 
on \tc{rotations of VLM objects}
set by theory. This is important and interesting for \tc{several} reasons. First, it is not difficult to imagine that this 
provides a hitherto unknown dependence of \mmhb with the (rapid) rotational speed $\Omega$, and as we show in sequel,
provides quantitative evidence for over-massive BDs, i.e., BDs with mass greater than \mmhb \tc{of the non-rotating case}. 
Second, for a given $\Omega$, we obtain a transition mass range \mmhb \tc{($\Omega$)} $\leq M \leq$ \mmax \tc{($\Omega$)} 
(with \mmax \tc{($\Omega$)} being a maximal mass) for stellar objects to evolve into main sequence stars.
Finally, we obtain an upper limit of the angular speed 
\tc{$\Omega_{\rm max} = 0.0047{\rm s}^{-1} $}, beyond which \tc{VLM objects do not evolve into main sequence stars}. 
\tc{Importantly, this is not the well known break up angular speed of a star which defines the limit for its disruption in 
Newtonian gravity via centrifugal forces. The latter is given in a standard fashion for a star of mass $M$ and 
radius $R$ by $(GM)^{1/2}/R^{3/2}$, where $G$ is the gravitational constant and one assumes spherical symmetry. This formula
is true irrespective of the stellar structure equations. To wit, a theoretically infinite number of $(M,R)$ tuples 
exist for which the Newtonian formula predicts a same break up angular speed. The numerical value quoted above
is on the other hand unique, and follows when we self-consistently solve the stellar structure equations in the presence
of rapid rotations. 
}

\tc{This last statement needs further clarifications. Indeed, as we have already said, the Newtonian formula
for the angular speed of disruption is only a quantification of mechanical equilibrium of the stellar object, and does not 
incorporate aspects of its thermal equilibrium. Specifically, given a limiting angular speed at which a stellar object 
of a given mass is disrupted, this formula gives an expression 
for its radius, which one can, to a crude approximation 
take as the volume equivalent radius of the object. It is however insensitive to the density distribution and 
degeneracy effects inside the stellar matter. Even if one assumes some form of these latter quantities, the Poisson's equation, which will
result in the gravitational potential at the surface of the object, will be difficult to solve as one does not
have well defined boundary conditions there. What we do on the other hand is to perform a self-consistent analysis 
in which mechanical equilibrium and thermal equilibrium are incorporated simultaneously.} 
 
\tc{To motivate the need for such analysis, consider, for example, the ratio of the hydrogen burning luminosity of the \tc{VLM object} 
to its surface luminosity. Physicality demands 
that this ratio is less than \tc{or equal to} unity. In a non-rotating scenario, if this ratio becomes slightly greater than one, then
the star will expand, in order to reduce the core temperature, thus achieving balance between internal nuclear energy
production and surface energy emission. \te{If the star is rapidly rotating with a given angular speed however, there can be a situation in which further reduction of core density due to internal readjustments, leads to non-existence of model solutions under the assumption of uniform rotation}. Clearly then, a consistent numerical analysis taking into account \te{mechanical equilibrium under} rapid rotation, in conjunction with the thermal equilibrium conditions, \tc{is required}.}

\tc{Here, we adopt the analytical polytropic model of \cite{BL}, with some parameters chosen according to Model D, of 
\cite{Chabrier1992} (the fourth row of Table 1 of that paper), 
via a numerical scheme to accommodate rapid rotation \te{(in the non-rotating case, this scenario has been used in 
\cite{Auddy}, \cite{ForbesLoeb}, \cite{BenitoWojnar})}.
Indeed, rotating polytropes have been studied for almost a century, starting with \cite{Jeans}, \cite{Chandra2}, and later by 
\cite{Roberts1}, \cite{Roberts2}, \cite{Roberts3}, \cite{James}, \cite{Stoeckly}, etc. The novelty of our work is the implementation
of the physics of \tc{VLM stars and} brown dwarfs in a rapidly rotating scenario.
In addition to the results mentioned above, we are also able to provide an analytical formula for the luminosity 
of a \tc{VLM object} when it reaches the main sequence, as a function of its mass and angular speed, within the ambits of this model. 
These are the main results of this paper. We 
should mention that we are in effect considering a toy analytical model of \tc{VLM objects}, where, 
apart from assuming a polytropic equation of state (EOS), atmospheric corrections are ignored. Modelling the atmospheres of
\tc{VLM stars} and BDs is indeed an active area of current research, see e.g. \cite{Marley}. Incorporating such models along with 
rapid rotations of \tc{VLM objects} is indeed a formidable challenge. Our simplified treatment on the other hand 
brings out several novel physical features of rapidly rotating VLM objects. 
}
  
The organisation of this paper is as follows. In the next section \ref{nonrot}, we recall
the basic features of non-rotating \tc{VLM objects}, and set up the analytical model. This is then used in section \ref{rot}
to include rotation, and we present our main results in the subsequent
section \ref{Results}. The paper ends with a summary in section \ref{Discussions}. 

\section{Non-rotating VLM objects}
\label{nonrot}

The basic assumptions that we use here are as follows. 
Firstly, the \tc{VLM object} is assumed to be fully convective, containing 
\tc{helium and partially ionised hydrogen mixture, with partially degenerate electrons in the interior, 
and helium and molecular hydrogen mixture at the photosphere}. The pressure, which arises due to 
both thermal effects (ions) and degeneracy (electrons) is considered to be non-relativistic. Importantly, 
these last two assumptions imply that we can safely use a polytropic EOS (polytropic
approximations are discussed in the textbook of \cite{Chandra}. For more discussions
on the applicability of this approximation to \tc{VLM objects}, see \cite{Poly2}, \cite{Poly1}). Further, it is assumed that 
the core temperature in \tc{VLM objects} is not sufficient to produce $He^4$. Hence the truncated $p-p$ chain thermonuclear 
reactions takes place in the stellar interior. Note that the original model of \cite{BL} model assumes 
spherical symmetry, which we will relax when we consider rapid stellar rotation. 

To set the stage, and to develop the notations used in the rest of this paper, we will now
recall some known facts in the evolution of \tc{VLM objects}. 
After its formation, during the initial stages, a \tc{VLM object} keeps contracting owing to its self gravity. In the process, 
it keeps radiating energy from its surface, which is referred to as surface luminosity $L_S$. 
Now, $L_S$ keeps decreasing as the \tc{object} contracts with time. The contraction, however, initially leads to an 
increase in its core temperature and density. It is known that rates of thermonuclear reactions are 
dependent on both of these. Hence, at some stage, if the attained core temperature and density are sufficient, 
then thermonuclear reactions start taking place. The energy generated within the \tc{object} due to this
is referred to as hydrogen burning luminosity $L_{HB}$. With further contraction of \tc{the object}, $L_{HB}$ 
starts increasing. A \tc{stellar object} is said to undergo stable/sustained hydrogen burning if the amount of energy 
liberated from surface is balanced by that produced from thermonuclear reactions within the star 
i.e., $L_S=L_{HB}$. Hence, at some point during the \tc{object}'s contracting phase, if stable hydrogen burning is attained, 
then further contraction ceases and the \tc{object} is said to become a main sequence star (MSS). However, if considerable 
amount of degeneracy sets in before the \tc{object} attains stability, 
a part of the thermal energy of the \tc{object} is used up in accommodating 
a large number of degenerate electrons in a smaller volume. This forbids the core temperature from rising further. 
The core temperature thus starts falling with further contraction. This eventually leads to a decrease in the 
$L_{HB}$ too, and hence the \tc{object} does not stabilise thermally. \tc{The object is then said to become a BD.} 

Now, if the initial mass of the \tc{object}, after formation, happens to be greater than a certain minimum value, 
then the \tc{object} eventually stabilises before the onset of considerable electron degeneracy. 
It is commonly believed that this minimum value, the $M_{\rm mmhb}$ (in the non-rotating case), 
sets the boundary between a MSS and a BD. Recently, however, \cite{ForbesLoeb} have shown that 
theoretically over-massive BDs (mass $\gtrsim$ $M_{\rm mmhb}$) 
are possible, via accretion effects. According to their analysis, $M_{\rm mmhb}$ should no longer demarcate between 
MSS and BD. However, it is still the minimum main sequence mass.

\begin{figure}[h!]
\centering
\includegraphics[width=0.34\linewidth]{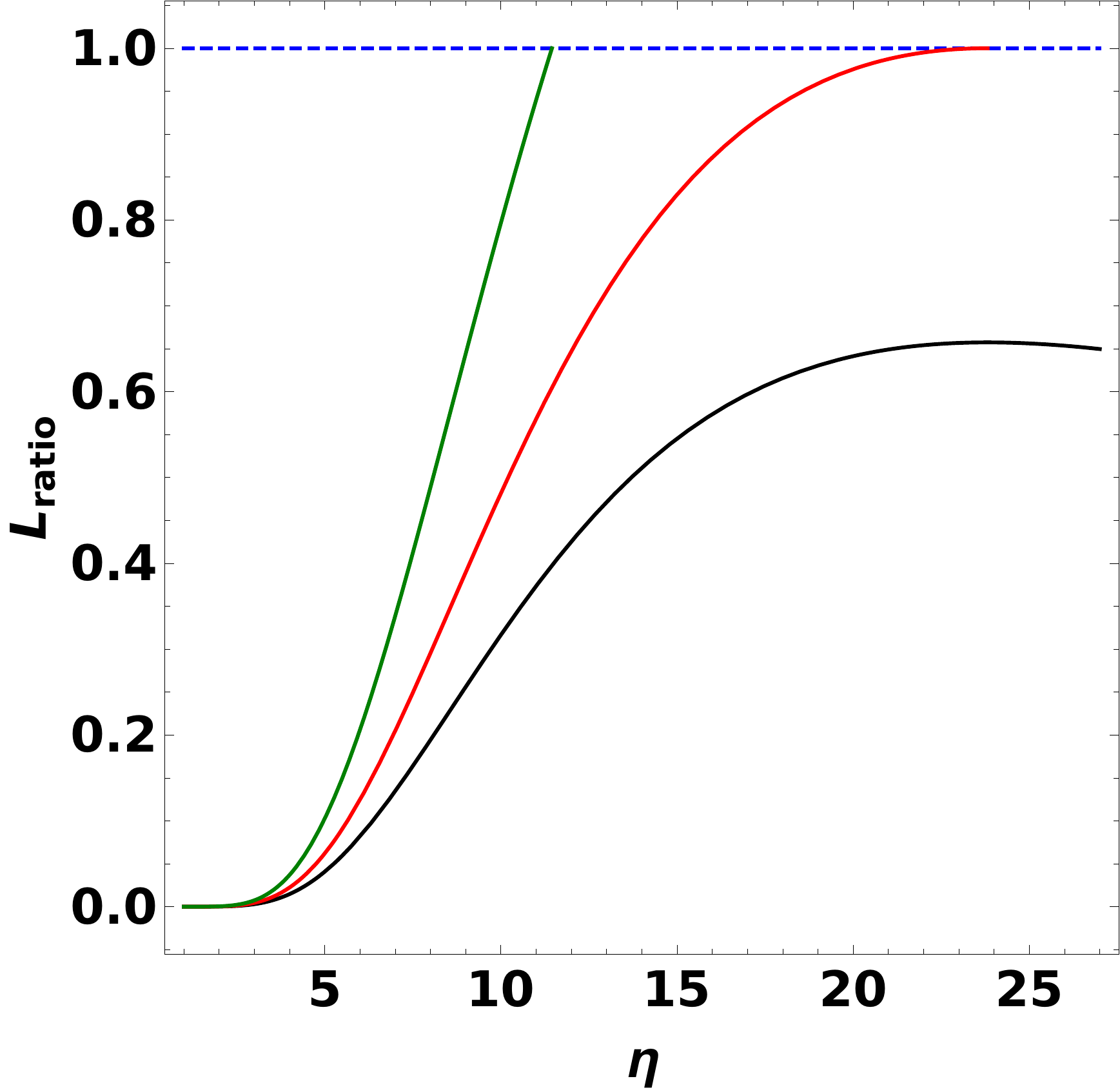}
\caption{\tc{$L_{\rm{ratio}}$ vs $\eta$ for non-rotating VLM objects : The red curve corresponds to $M_{\rm mmhb}$($0.081M_{\odot}$), 
	the green one for mass $=0.085M_{\odot}$ and the black one for $0.078M_{\odot}$.}}
\label{Lratio0}
\end{figure}

The continuous contraction of a \tc{VLM object}, after its formation, leads to increase in its degeneracy. 
Hence one simulates the time-evolution of \tc{an object} of a given mass, by varying the degeneracy parameter 
(called $\eta$ in sequel). 
For an \tc{object} with a given mass, we compute $L_{HB}$ and $L_S$ at every instant of its contracting phase 
(i.e., lower $\eta$ value to higher ones). In the process, for every $\eta$, we get the value of the ratio 
of the two luminosities, $L_{\rm{ratio}} =\frac{L_{HB}}{L_S}$. We then plot $L_{\rm{ratio}}$ vs $\eta$ for the 
given mass. From the above discussions, we see that $L_{\rm{ratio}}$ first increases, and then starts descending 
after attaining a maxima. If the maximum value is less than unity, then it indicates that the \tc{object} can 
never reach the stable hydrogen burning condition $L_{HB}=L_S$. Hence, we repeat the above numerical procedure 
for a mass higher than the one previously chosen and repeat the numerical procedure, 
till the maxima of the plot attains unity. This mass is then the $M_{\rm mmhb}$.
At the point where \tc{an object} of a given mass attains stable hydrogen burning (i.e. $L_{\rm{ratio}}=1$), it has evolved into 
a MSS. \tc{Thus, from that point onwards one needs to consider 
a MSS model to further track the evolutionary process.} The discussion above is illustrated in
Fig. \ref{Lratio0}, from which one can see that \tc{objects} having 
masses greater than $M_{\rm mmhb}$ attains stability (i.e. $L_{\rm{ratio}}=1$) at lower $\eta$ 
values. 

\section{Effects of rotation in the VLM objects' evolutionary process}
\label{rot}

We first consider the \tc{VLM object} to be centered at the origin of a Cartesian coordinate system \{$x^1,x^2,x^3$\}. 
Now we consider uniform rotation of the \tc{object} along the $x^2$-axis. Owing to centrifugal forces, the \tc{object} bulges near the 
equatorial plane (i.e. $x^2=0$ plane), deforming it into an oblate spheroid. Hence, \tc{an object} under rapid rotation 
loses spherical symmetry. \tc{The essential features of the analytic model, chosen for studying effects of 
rapid rotation, are the same as that of \cite{BL}}, excepting for the assumption of spherical symmetry there, which
will lead to crucial modifications as we discuss below. 

\subsection{The stellar equations and numerical recipe }
\label{stellarEqnNumRecipe}

We begin with the polytropic equation of state, the Poisson's equation, and the Euler equation corresponding to momentum conservation.
The polytropic equation reads
\begin{equation}
\label{polytopric}
P=\kappa\rho^{(1+\frac{1}{n})}~,
\end{equation}
where $P$ is the pressure, $\kappa$ is the polytropic constant and $n$ the polytropic index, which is to be taken as $1.5$,
as appropriate for \tc{VLM objects}. The Poisson's equation reads
\begin{equation}
\label{poisson}
\nabla^2 \phi=4 \pi G \rho~,
\end{equation}
where $\rho$ is the density and $\phi$ is the gravitational potential.
Finally, the Euler equation corresponding to momentum conservation is
\begin{equation}
\label{Euler}
\rho \frac{\partial v^i}{\partial t} + \rho v^j \frac{\partial v^i}{\partial x^j} = 
-\frac{\partial P}{\partial x^i} - \rho \frac{\partial \phi}{\partial x^i}~,
\end{equation}
where $t$ is the temporal coordinate and $v^i=\Omega\{x^3,0,-x^1\}$ is the velocity field of the \tc{object}, 
with $\Omega$ being its uniform angular speed.

We get the deformed equilibrium configuration of the rotating \tc{object} of a fixed mass $M$, by numerically solving Eqs. (\ref{poisson}) 
and (\ref{Euler}), for a given $\kappa$. Initially we solve the Lane-Emden equation 
to obtain the spherically symmetric density profile, 
which is then used in Eq. (\ref{poisson}), to obtain the gravitational potential $\phi$. Then using 
this $\phi$ in Eq. (\ref{Euler}), we obtain the updated density profile $\rho$. We feed this updated 
$\rho$ back into Eq. (\ref{poisson}), to obtain an updated $\phi$, which in turn yields an updated 
$\rho$ from Eq. (\ref{Euler}). This iteration is repeated until a desired convergence is achieved for a given 
tuple \{$M,\Omega,\kappa$\}. 
For further details of the numerical procedure, the reader is referred to 
\cite{Ishii}, \cite{tapo3}.

From the converged solution, we also obtain the central density $\rho_c$ of the deformed \tc{object} in equilibrium.
\tc{Now, the polytropic constant is related to the degeneracy parameter ($\eta$) of the object as follows}
\begin{equation}
\label{kappaeta}
\kappa=\frac{(3 \pi^2)^{\frac{2}{3}}\hbar^2}{5 m_e m^{\frac{5}{3}}_H \mu^{\frac{5}{3}}_e}
\Big(1+\frac{\alpha}{\eta}\Big), ~~~~~~~
\eta=\frac{(3 \pi^2)^{\frac{2}{3}}\hbar^2}{2 m_e m^{\frac{2}{3}}_H\mu^{\frac{2}{3}}_e k_B}\frac{\rho^{\frac{2}{3}}}{T}~,
\end{equation}
where $m_e$ and $m_H$ denote the electron and hydrogen mass respectively, and \tc{$\alpha=5\mu_{e}/2\mu_{1}$}. \tc{Here, $\mu_e$ is 
the number of baryons per electron [$\mu_e=1.143$] and $\mu_{1}$ is the mean molecular weight of the helium and the partially 
ionised hydrogen mixture in the interior [$\mu_{1}=0.996$ for Model D of \cite{Chabrier1992}, see Table 1 of \cite{Auddy}]}. 

Also note that in Eq. (\ref{kappaeta}), $\eta$ is defined to be the ratio of the Fermi energy to $k_B T$, 
where $T$ is the temperature, $k_B$ being the Boltzmann constant. 
Thus fixing $\eta$ inherently determines the corresponding $\kappa$.
We then compute the central temperature $T_c$ from Eq. (\ref{kappaeta}), using the obtained value of $\rho_c$ for 
the converged equilibrium configuration of the deformed \tc{object}. 
Using these, we numerically calculate the hydrogen burning luminosity of the \tc{object},
\begin{equation}
\label{lhb}
L_{HB}= \int_V \rho\epsilon \,dx^1\,dx^2\,dx^3~,
\end{equation}
where $\rho$ and $\epsilon$ are functions of the spatial coordinates \{$x^1,x^2,x^3$\}, 
with $\epsilon$ being the energy generation rate per unit mass. The integration is performed numerically over 
the deformed volume $V$. For \tc{VLM objects}, \te{with typical values of the core temperature $T_c \sim 3\times 10^6$ K
and core density $\rho_c \sim 10^3 gm~cm^{-3}$, we can fit $\epsilon$ with a power law in $T$ and $\rho$ following
\cite{BL}, and obtain} 
\begin{equation}
\label{energygen}
\epsilon=\epsilon_c\Big(\frac{T}{T_c}\Big)^s\Big(\frac{\rho}{\rho_c}\Big)^{u-1},~~
\epsilon_c=\epsilon_0 T_c^s\rho_c^{u-1}~ergs~g^{-1}s^{-1}
\end{equation}
with $s \simeq 6.31$, $u\approx2.28$ and $\epsilon_0=1.66\times10^{-46}$.
Next, we compute the luminosity at the photosphere (i.e., surface luminosity) of the deformed \tc{object}. 
At any point near the surface, the atmosphere can be locally approximated to be plane parallel, irrespective 
of rotation. Thus we use the following definition of optical depth ($\tau(z)$) for a planar atmosphere 
to determine the location of photosphere in a deformed \tc{object} :
\begin{equation}
\tau(z)=\int_{z}^{\infty} \kappa_{R} \rho dz~,
\end{equation}
where $z$ is the local vertical depth of the atmosphere and $\kappa_{R}$ is the Rosseland 
mean opacity, \te{taken here to be $0.01~cm^2~g^{-1}$, which is an order of magnitude estimate, 
being roughly a tenth of the free electron opacity (see \cite{BL}, \cite{ForbesLoeb}).}
The photosphere is then defined to be located at $z_e$ for which $\tau(z_e)=2/3$. 
The temperature $T_e$ and the density $\rho_e$ at the photosphere are related through 
\tc{\begin{equation}
\label{photoTrho}
\frac{T_e}{K}=\frac{b_{1}\times10^6}{\eta^{\nu}}\Big(\frac{\rho_e}{g/cm^3}\Big)^{0.4}
\end{equation}
where $b_{1}=2.0$ and $\nu=1.60$ for Model D of \cite{Chabrier1992}, see again Table 1 of \cite{Auddy}.}
Now using Eq. (\ref{photoTrho}) and the ideal gas law, and assuming approximate constancy of the acceleration due to gravity
near the surface, we obtain a local expression for the temperature at the photosphere.
\tc{\begin{equation}
T_e=\Big(\frac{2 g \mu_{2} m_H}{3 \kappa_{R} k_B}\Big)^{\frac{0.4}{1.4}}\Big(\frac{b_{1}\times 10^6}
{\eta^{\nu}}\Big)^{\frac{1}{1.4}}~.
\end{equation}
where $\mu_{2}$ is the mean molecular weight of the helium and molecular hydrogen mixture at the photosphere [$\mu_{2}=2.285$].} 
For a deformed \tc{object}, the relative position of the photosphere with respect to the surface of the \tc{object} does not remain 
constant throughout, i.e., it varies from one surface point to another, unlike the case for a spherically symmetric 
\tc{object}. This would also be the case with $T_e$. Now applying Stefan-Boltzmann law, we compute the total surface luminosity as
\begin{equation}
\label{ls}
L_S=\oint_S \sigma {T_e}^4 \,dA
\end{equation}
where $dA$ is the elemental surface area and $\sigma$ is the Stefan-Boltzmann constant. 
The integration is performed over the entire surface of the deformed \tc{object}.
Finally, we compute the $L_{\rm{ratio}}$ for the given \{$M, \Omega,\eta$\}, which would be of fundamental 
importance to decide upon the fate of a rotating \tc{object}'s evolution. It should be noted that
in the limiting case of $\Omega \to 0$, we recover the original model due to \cite{BL}. 

\section{Results and analysis}
\label{Results}
We now present the main results obtained from our computational scheme discussed above. 

\subsection{$M_{\rm mmhb}$ as a function of $\Omega$}

\tc{Rotation tends to reduce the strength of gravity inside a star. This effectively makes a rotating star, achieve hydrostatic equilibrium at a lower core temperature and density. As a result total nuclear energy production is reduced, so a higher mass is required to achieve thermal stability ($L_{\rm ratio}=1$).} Hence we find that the $M_{\rm mmhb}$, in presence of rotation to be larger than non rotating case.
In order to find $M_{\rm mmhb}$ corresponding to a given stellar rotation $\Omega$, we carry out a similar algorithm 
described in section \ref{nonrot}, using the numerical prescription mentioned in section \ref{stellarEqnNumRecipe}. 
We perform this numerical procedure for different $\Omega$ values, to obtain a fitted formula 
for $M_{\rm mmhb}$ as a function of $\Omega$. We find,
\begin{equation}
M_{\rm mmhb}(\Omega)= 0.0814 + 0.2302 \Omega + 245.58 \Omega^2 + 67646 \Omega^3~,
\label{MMHBformula}
\end{equation}
where $\Omega$ is in $s^{-1}$ and the formula gives $M_{\rm mmhb}(\Omega)$ in units of the solar mass $M_{\odot}$. 
From Eq. (\ref{MMHBformula}), we find that \mmhb increases monotonically with $\Omega$ as depicted by the red
curve in Fig. \ref{MMHB_MMAX}, 
but that for small $\Omega$, the change from the case $\Omega = 0$ is maximally
by a few percents. For example, for the smallest observed period of $1.08$ h of \cite{Tannock}, the increase in
\mmhb is by $\sim 1.6\%$. Faster rotations can however significantly change the result. Using the period of $17$ min 
of \cite{Route}, the increase in \mmhb is $\sim 33\%$. However, such a small value of the period is ruled
out of our analysis, as we momentarily see. We will find that the minimum period of a \tc{VLM star} can be \tc{$\sim 22$ min}, 
and hence the maximal increase in \mmhb is $\sim 17\%$. Importantly, we have demonstrated that 
rotation can lead to the existence of \tc{over-massive BDs without accretion effects}.

\te{As an aside, we note the behavior of \mmhb with $\Omega$ for the two extreme models of \cite{Chabrier1992} - Model A and Model H in \cite{Auddy}. For Model A we obtain \mmhb$(\Omega) = 0.0879 + 0.2441 \Omega + 220.53 \Omega^2 + 60446 \Omega^3 $, while for Model H we obtain \mmhb$(\Omega) = 0.0636 + 0.2113 \Omega + 445.97 \Omega^2 + 117094 \Omega^3$. The increase in \mmhb, corresponding to the smallest observed rotation period of $1.08$ h of \cite{Tannock}, is by $\sim 1.4\%$ for Model A and $\sim 3.1\%$ for Model H. The smaller period of $17$ min of \cite{Route} corresponds to an increase in \mmhb by $\sim 27\%$ and $\sim 72\%$ for Model A and Model H respectively. However, such a small value of the rotation period $\sim 17$ min, falls below the minimum permissible periods corresponding to both Model A and H, and is thus ruled out of our analysis. The minimum permissible period for Model A is $\sim 20$ min, which corresponds to a maximal increase in \mmhb by $\sim 19\%$. For Model H, the maximal increase in \mmhb is by $\sim 21\%$ corresponding to the minimum permissible period of $\sim 28$ min.}

\subsection{Behavior of $L_{\rm ratio}$ vs $\eta$ plot for non-zero stellar rotation :}
\begin{figure}[h!]
	\centering
	\hspace{0.3cm}
	\begin{minipage}[t]{0.47\linewidth}
		\begin{center}
			\centerline{\includegraphics[scale=0.35]{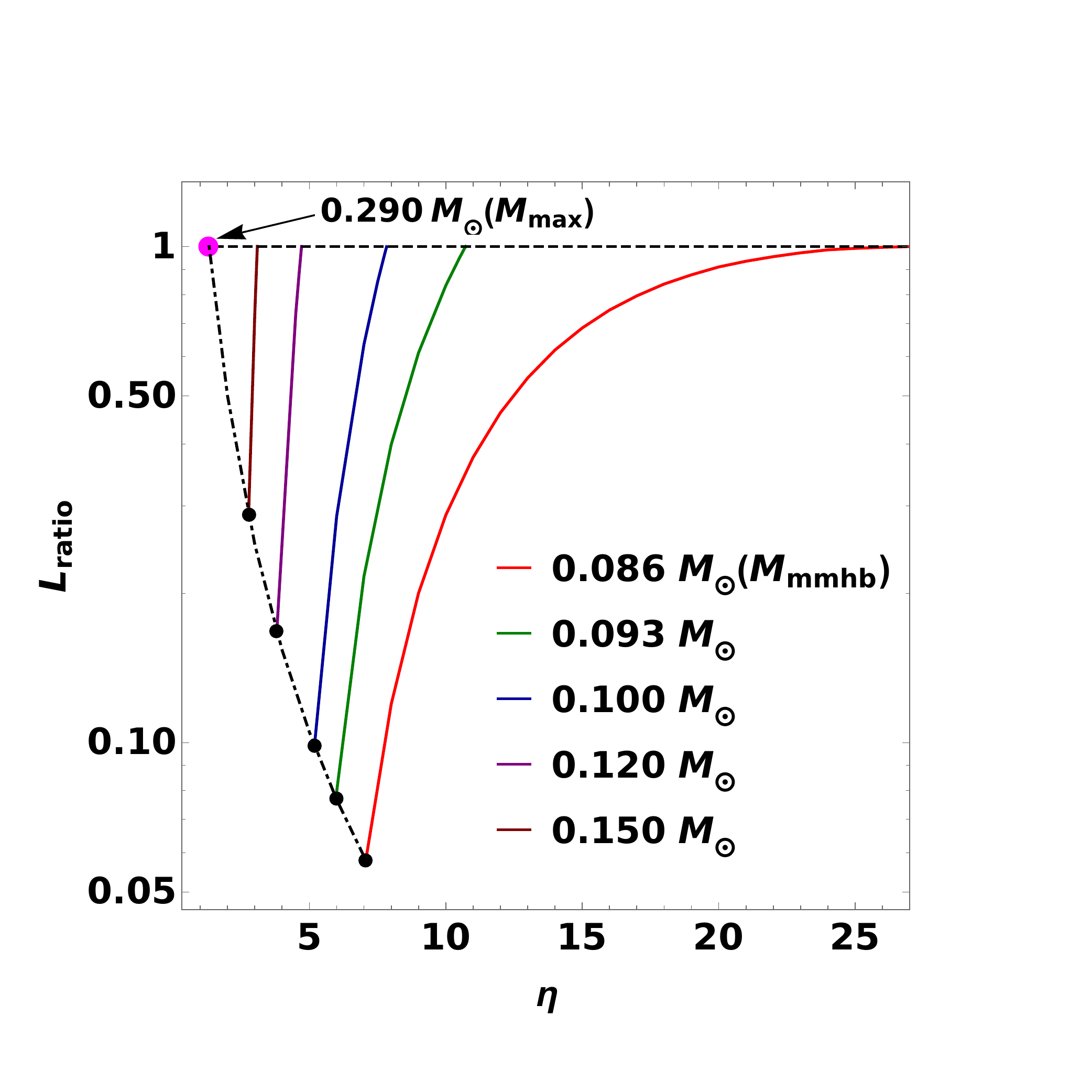}}
			\caption{\tc{$L_{\rm ratio}~ vs~ \eta$ plot for $\Omega=0.003s^{-1}$. The red curve corresponds to $M_{\rm mmhb}$. 
			As explained in the text, it starts from $\eta=\eta_{\rm crit}$ and non-zero critical $L_{\rm ratio}$, represented by the corresponding
			filled black circle. 
It reaches main sequence when $L_{\rm ratio}=1$. For higher masses, $\eta_{\rm crit}$ is lower and critical $L_{\rm ratio}$ is higher. Higher mass objects reach the main sequence at lower values of $\eta$. The mass value for which critical $L_{\rm ratio}=1$ (here $0.29 ~ M_{\odot}$), is the $M_{\rm max}$ for this
$\Omega$. The black dot-dashed line represents the locus of critical points (marked by filled circles) of the curves corresponding to different masses ranging from $M_{\rm mmhb}$ to $M_{\rm max}$, for this $\Omega$. $M_{\rm max}$ is marked by the filled magenta circle.}}
			\label{Lratio_Eta_omega0p003}
		\end{center}
	\end{minipage}%
	\hspace{0.3cm}
	\begin{minipage}[t]{0.47\linewidth}
		\begin{center}
			\centerline{\includegraphics[scale=0.48]{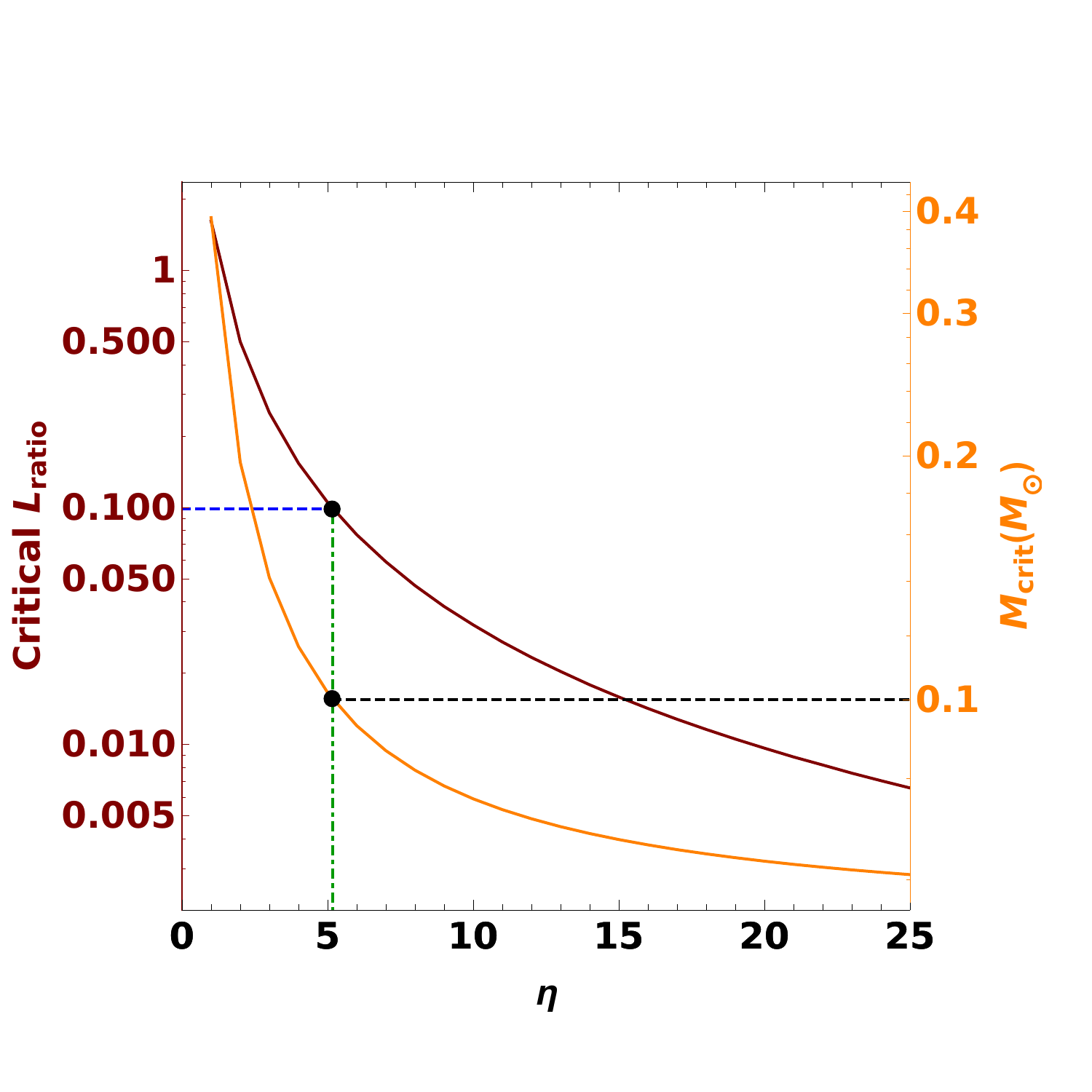}}
			\caption{\tc{Critical $L_{\rm ratio}~vs~\eta$ and $M_{\rm crit}~vs~\eta$ plot for $\Omega=0.003s^{-1}$. The brown curve corresponds to critical $L_{\rm{ratio}}$, while the orange one corresponds to $M_{\rm{crit}}$. The point on the critical $L_{\rm{ratio}}$ curve, which corresponds to $0.098$, is the top filled black circle. The vertical line from that point intersects the $\eta$ axis at corresponding $\eta_{\rm{crit}}(5.18)$. The point of intersection between the same vertical line and the $M_{\rm{crit}}$ curve, gives the corresponding $M_{\rm{crit}}(0.1M_{\odot})$.
				Each y-axis is color coded to the data.}}
			\label{critPlot}
		\end{center}
	\end{minipage}
\end{figure}

\tc{
The $L_{\rm ratio}$ vs $\eta$ plot for VLM objects corresponding to a given non-zero rotation $\Omega$ is very different from 
its non-rotating counterpart (compare Fig. \ref{Lratio0} with Fig. \ref{Lratio_Eta_omega0p003}).}
\tc{Here we observe that for a given mass and rotation, the $L_{\rm ratio}$ vs $\eta$ curve starts from a particular point (let us call it the critical point). The particular values of $\eta$ and $L_{\rm ratio}$ corresponding to the critical point are referred to as the $\eta_{\rm crit}$ and critical $L_{\rm ratio}$ respectively, for that particular mass and rotation. As an example, 
$\eta_{\rm crit}=5.18$ and critical $L_{\rm ratio}=0.09$ for a VLM object of mass $M=0.1M_{\odot}$ rotating with angular speed of $\Omega=0.003s^{-1}$ 
(the filled black circle where the blue line culminates in Fig. \ref{Lratio_Eta_omega0p003}). A VLM object of given mass and rotation, cannot achieve hydrostatic equilibrium for $\eta$ values lower than its corresponding $\eta_{\rm crit}$. The reason is as follows. A VLM object cannot sustain a given rotation, unless it possesses a certain minimum central density, dubbed as critical density, $\rho_{\rm{crit}}$. We also know that during the contraction phase of a VLM object, its central density $\rho_c$ increases with an increase in degeneracy $\eta$. Hence the central density at $\eta_{\rm crit}$ for any VLM object of a given mass and rotation, denotes the minimum value of central density below which the object cannot sustain the applied rotation (i.e., for central densities corresponding to $\eta\leq\eta_{\rm crit}$ \te{no model solution exists}). As Fig. \ref{Lratio_Eta_omega0p003} indicates, 
for a particular $\Omega$, higher mass VLM objects attain the corresponding critical density at lower $\eta_{\rm crit}$ values 
and the corresponding critical $L_{\rm ratio}$ are higher. It is observed from that figure 
that for a given rotation, higher mass objects reach the main sequence at lower value of $\eta$. 
The mass value for which critical $L_{\rm ratio}=1$ will be called as ${M_{\rm max}}$ for the given $\Omega$. For example $M_{\rm max}=0.29 ~ M_{\odot}$ 
for $\Omega=0.003s^{-1}$ (See Fig. \ref{Lratio_Eta_omega0p003}). In Fig. \ref{Lratio_Eta_omega0p003}, the black 
dot-dashed line represents the locus of critical points (marked by filled circles) of the curves corresponding to different masses 
ranging from $M_{\rm mmhb}$ to $M_{\rm max}$ and is a portion of the critical $L_{\rm ratio}~vs~\eta$ curve 
in Fig. \ref{critPlot}, which we now explain.}

\subsection{The Existence of $M_{\rm max}$}
\label{existenceMmax}
\tc{ 
As we have already indicated, for central densities lesser than $\rho_{\rm{crit}}$, \te{no model solution is possible for a VLM object of given mass and angular speed for which we are referring the $\rho_{\rm{crit}}$.} We call the stellar configuration as a critical configuration when the 
central density equals $\rho_{\rm{crit}}$.
\tc{For a given $\Omega$, we find the critical configuration for each value of the degeneracy $\eta$. 
For each of these critical stellar configurations, 
we record the critical density $\rho_{\rm{crit}}$ and compute the critical value of $L_{\rm{ratio}}$ as well as
the critical mass, which we call $M_{\rm{crit}}$. While we find that $\rho_{\rm{crit}}$ remains constant with $\eta$, 
Fig. \ref{critPlot} indicates that both critical $L_{\rm{ratio}}$ and $M_{\rm{crit}}$ falls with increasing degeneracy, as
is not difficult to justify physically from the following :\\
\begin{itemize}
	\item For a given $\Omega$, we have seen that the critical density remains constant with increase in $\eta$. We also know with increase in $\eta$, the $\kappa$ value decreases according to Eq. (\ref{kappaeta}). Hence, with increase in $\eta$ the central pressure for the corresponding critical configurations keeps decreasing according to the formula for polytropic EOS. We know that reduced central pressure can support lower stellar mass. Hence, for a given $\Omega$, increase in $\eta$ leads to decrease in $M_{\rm crit}$.
	\item For a given $\Omega$, with increase in $\eta$ value, the central temperature corresponding to the critical configuration keeps decreasing according to Eq. (\ref{kappaeta}), owing to constancy of $\rho_{\rm crit}$. This leads to reduction in $L_{HB}$ and hence $L_{\rm ratio}$ for the corresponding critical configurations.
\end{itemize}
The information that one obtains from the plot in Fig. \ref{critPlot} is the particular minimum value of $\eta~(=\eta_{\rm crit})$ at which a VLM object of given mass $M_{\rm crit}$, has just the sufficient central density $\rho_{\rm{crit}}$, in order to sustain the given rotation $\Omega$. One also obtains the value of the corresponding critical $L_{\rm ratio}$. From $\eta_{\rm crit}$ onwards, as the object of that particular mass $M_{\rm crit}$ keeps contracting, the systematic behavior of $L_{\rm ratio}$ vs $\eta$ follows. At this point one can draw complete correspondence between Fig. \ref{critPlot} and Fig. \ref{Lratio_Eta_omega0p003}. For example, in Fig. \ref{critPlot}, we see a VLM object of mass $0.1M_{\odot}$ attains the critical density at $\eta_{\rm crit}=5.18$ and the corresponding critical $L_{\rm ratio}=0.098$. This particular mass object \te{will not possess any model solution for $\eta$ values less than $\eta_{\rm crit}$ under the given rotation}. Hence, we see from Fig. \ref{Lratio_Eta_omega0p003}, the $L_{\rm ratio}$ vs $\eta$ curve for object of mass $0.1M_{\odot}$ starts from the point ($\eta=5.18$, $L_{\rm ratio}=0.098$). From that point onwards, as the given object contracts, the central density keeps increasing along with an increase in degeneracy $\eta$. Thus the \te{model solutions for the object will exist} under the given rotation for $\eta\geq\eta_{\rm crit}$, and we get the systematic plot of $L_{\rm ratio}$ vs $\eta$ for that particular object starting from $\eta=\eta_{\rm crit}$.}

\tc{For a given $\Omega$, the critical $L_{\rm{ratio}}$ value corresponding to $\eta_{\rm{crit}}$ of a specified mass object reveals the relative 
magnitude of hydrogen burning luminosity and surface luminosity, at the initial stage of its evolution, when it has attained sufficient
degeneracy to sustain the given rotation. Three situations can arise at this particular critical point. 
Firstly, if critical $L_{\rm{ratio}}<1.0$, the object will attain stability with further evolution, 
only if $M \geq M_{\rm mmhb}(\Omega)$. Secondly, for critical $L_{\rm{ratio}}=1$, the object has already turned into a MSS. We label the corresponding mass as $M_{\rm max}$ for the given $\Omega$. Finally, for critical $L_{\rm{ratio}}>1.0$, the hydrogen burning luminosity exceeds surface luminosity and our VLM object can never stabilize \te{while maintaining the given uniform rotation.} We will show this explicitly in a moment. 
Hence, for a given $\Omega$, the valid range of mass $M$, for which a VLM object can eventually 
evolve into a MSS, is $M_{\rm mmhb}\leq M \leq M_{\rm max}$. We call this $M_{\rm max}$ as the maximum mass of stable 
hydrogen burning for this $\Omega$ and as is clear from the context, the quantity ${M_{\rm max}}$ is
purely an artefact of rotational effects. We shall refer to this mass range as the transition mass range.}

Let us now comment upon the case critical $L_{\rm{ratio}}>1.0$. To understand this, we use the standard stellar energy equation
${\dot\epsilon} - \partial L/\partial M = T dS/dt$, where $L$ denotes the luminosity, $M$ the mass, $T$ the temperature 
and $S$ is the entropy per unit mass. Integrating this equation, we get after a little bit of algebra, 
\tc{\begin{equation}
L_S(1-L_{\rm ratio}) = \frac{K}{\eta^2}\frac{d\eta}{dt}\int\rho^{5/3}dV~,~~K = 5.25\times 10^5 \frac{N_Ak_B}{\mu_e^{2/3}}~,
\label{Lratioeq}
\end{equation}}
with $N_A$ being Avogadro's number and $k_B$ the Boltzmann's constant. 
Clearly then, at $\eta_{\rm crit}$, if critical $L_{\rm ratio}>1$, Eq. (\ref{Lratioeq}) dictates that $d\eta/dt <0$, and in this case, the 
object \te{will not possess a model solution below the minimum degeneracy $\eta_{\rm crit}$ for the given $\Omega$.}

\subsection{$M_{\rm max}$ as a function of $\Omega$}

The concept of a critical density is only valid for a rotating stellar \tc{object}. 
For a non-rotating stellar object, $M_{\rm max}$ is not defined. 
This means that any non-rotating \tc{object} of mass $M \geq M_{\rm mmhb}$ can in principle evolve to the main sequence. 
\tc{In order to find $M_{\rm max}$ corresponding to a given non-zero stellar rotation $\Omega$, we numerically 
compute the $M_{\rm{crit}}$ value corresponding to critical $L_{\rm ratio}=1.0$. For example from Fig. \ref{critPlot}, one can see that the $M_{\rm crit}$ value corresponding to critical $L_{\rm ratio}=1.0$ is $0.29M_{\odot}$, which is the $M_{\rm max}$ for $\Omega=0.003s^{-1}$.} We perform this numerical procedure for different $\Omega$ values, 
to obtain a fitted formula for $M_{\rm max}$ as a function of $\Omega$. We find in units of $M_{\odot}$,
\tc{\begin{eqnarray}
M_{\rm max}(\Omega) =  -1.2621 + 0.0176\Omega^{-1} - 7.7873 \times 10^{-5}\Omega^{-2} + 1.1648 \times 10^{-7}\Omega^{-3}
\label{MMAXformula}
\end{eqnarray}}
\tc{where $\Omega$ is in $s^{-1}$.} 
\begin{figure}[h!]
	\centering
	\includegraphics[width=0.4\linewidth]{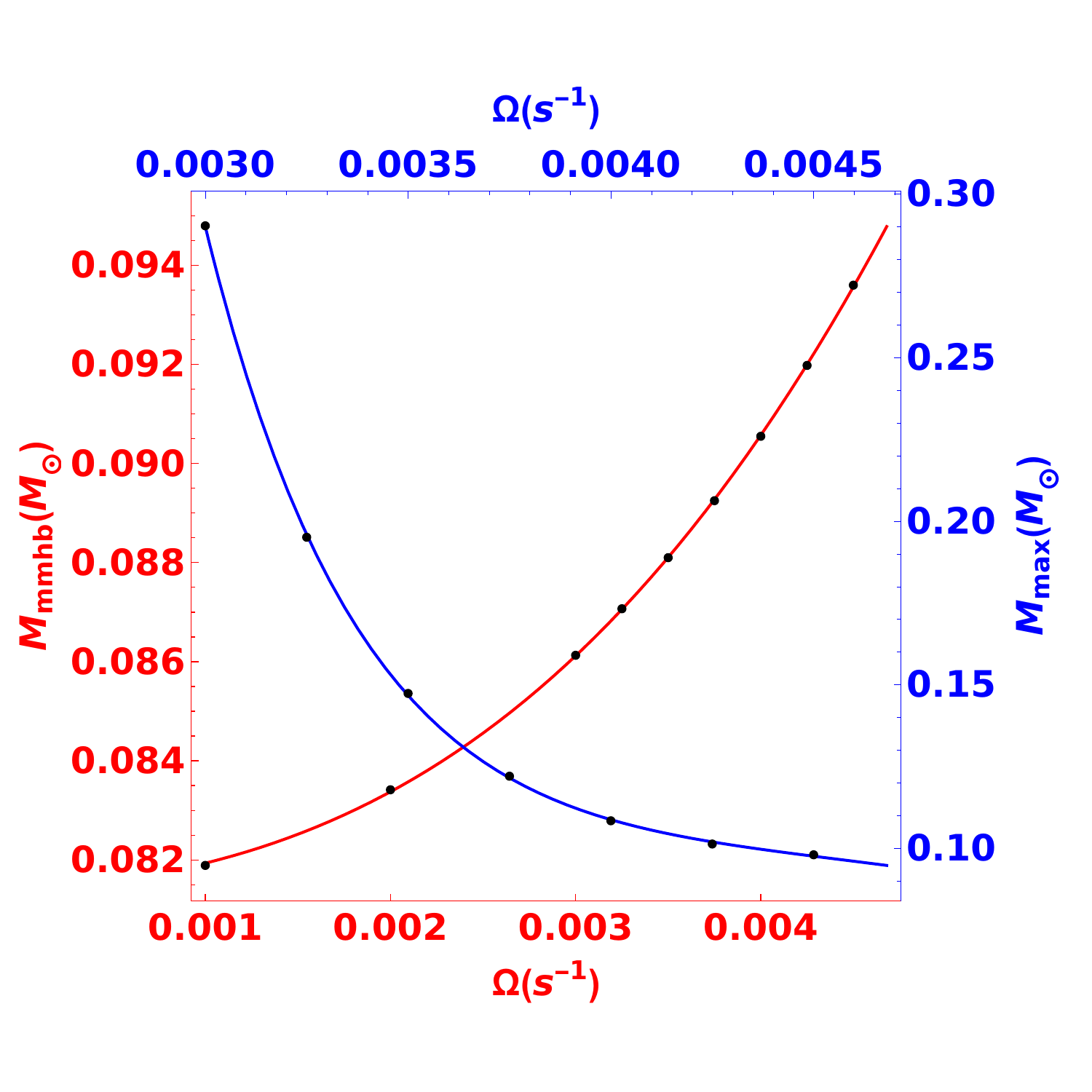}
	\caption{\tc{The red curve corresponds to $M_{\rm mmhb}(\Omega)$, while the blue one corresponds to $M_{\rm max}(\Omega)$. 
	Each y-axis is color coded to the data.}}
	\label{MMHB_MMAX}
\end{figure}
From Fig. \ref{MMHB_MMAX}, which depicts this behavior, one can see that with increase in $\Omega$, 
$M_{\rm max}$ decreases. \tc{This can be explained as follows. We already know that for a given $\Omega$, $M_{\rm max}$ corresponds to mass of the particular critical configuration, for which stable hydrogen burning takes place at the critical point. Now, for a higher value of $\Omega$, the corresponding critical configuration maintains hydrostatic equilibrium at higher central density and temperature. As a consequence, total nuclear energy production gets magnified, resulting in critical $L_{\rm ratio} > 1.0$. So a lower mass is needed to attain thermal stability at the critical point (i.e. critical $L_{\rm ratio}=1.0$).} As a consequence, \mmax decreases with increase in $\Omega$.

\tc{For $\Omega$ values less than $0.003s^{-1}$, the corresponding $M_{\rm max}$ values are larger than $0.3M_{\odot}$ beyond which an object is no longer in fully convective equilibrium. Hence those points are not shown in Fig. \ref{MMHB_MMAX}.}
 
\tc{From the behavior of $M_{\rm mmhb}$ and $M_{\rm max}$ with $\Omega$, we see gradual decrease in the transition mass range $[M_{\rm mmhb}(\Omega),M_{\rm max}(\Omega)]$, with increase in $\Omega$. There exists a certain $\Omega=0.0047~s^{-1}$, where the two curves $M_{\rm mmhb}(\Omega)$ and $M_{\rm max}(\Omega)$ meet (this is not the one obtained by visual inspection in Fig. \ref{MMHB_MMAX} where the two quantities are drawn with different scales). Consequently the distinctive transition mass range reduces to a single point at this particular $\Omega$. We shall call this angular speed $\Omega_{\rm max}$. Thus according to our model, for stellar rotations with angular speeds more than $\Omega_{\rm max} \sim 0.0047~s^{-1}$ (or rotation periods less than $\sim 22$ min), a VLM object cannot evolve into a MSS.}

\subsection{Stable Luminosity Formula}

Finally, we deduce a formula for the stellar Luminosity ($L_{HB}$ due to H-burning), at the point 
when \tc{object} reaches the main sequence, after initial evolution. This is denoted by $\tilde{L}_{HB}$ and is 
a function of both the stellar mass 
and the stellar rotation. The lowest order polynomial which best fits our generated data is represented
as $\tilde{L}_{HB}(M,\Omega)/L_{\odot} = \sum_{\alpha,\beta}C_{\alpha\beta}(M/M_{\odot})^{\alpha}(\Omega/s^{-1})^{\beta}$, and 
the coefficients $C_{\alpha\beta}$ are listed in Table \ref{TableL}. 
\begin{table}[h!]
	\centering
	\renewcommand{\arraystretch}{1.3}
	\caption{List of coefficients $C_{\alpha\beta}$}
	\label{TableL}
	\resizebox{\columnwidth}{!}{\begin{tabular}{|c| c| c| c| c| c| c| c| }\hline
			$\alpha \backslash \beta$ &~0 &1 &2 &3 &4 &5 &6 \\ \hline
			$0$& $24.203$& $1265.4$& $95238$& $1.2927\times 10^{6}$& $6.7610\times10^6$& $7.3972\times10^9$& $1.114\times10^{11}$ \\
			$1$& $-1577.2$& $-70298$& $-4.1388\times10^6$& $-3.3679\times10^7$& $-8.8127\times10^8$& $-1.0444\times10^{11}$& $-$\\
			$2$& $42761$& $1.5598\times10^6$& $6.6859\times10^7$& $2.8294\times10^8$& $1.0488\times10^{10}$& $-$& $-$ \\
			$3$& $-6.1736\times10^5$& $-1.7279\times10^7$& $-4.7498\times10^8$& $-7.9453\times10^8$& $-$& $-$& $-$ \\
			$4$& $5.0065\times10^6$& $9.5552\times10^7$& $1.2490\times10^9$& $-$& $-$& $-$& $-$ \\
			$5$& $-2.1626\times10^7$& $-2.1102\times10^8$& $-$& $-$& $-$& $-$& $-$ \\
			$6$& $3.8887\times10^7$& $-$& $-$& $-$& $-$& $-$& $-$ \\
			\hline
	\end{tabular}}
\end{table}
Fig. \ref{Lhbcontours} represents the contour plot of $\tilde{L}_{HB}$. All the \tc{objects}, 
having particular masses and rotations ($M,\Omega$) tuples, constituting any given contour, 
will end up in the main sequence, with the same luminosity. The results from this plot are not 
to be extrapolated beyond the valid range of \tc{transition mass} ($M_{\rm mmhb}(\Omega) \leq M \leq M_{\rm max}(\Omega)$, 
where $\Omega \in (0,\Omega_{\rm max})$), since beyond this range, \tc{an object} never evolves into a MSS.

\tc{It has been observed that our model parameters (corresponding to Model D of \cite{Chabrier1992}) succeeds in reproducing reasonable estimates of stellar luminosities, characteristic to the VLM objects up to a maximal mass of $\sim 0.1M_{\odot}$. Hence the confidence in our model lies precisely in the region:\\
\begin{eqnarray}
M_{\rm mmhb}(\Omega)\leq M\leq M_{\rm max}(\Omega)~,~~{\rm with}~~ M \leq 0.1M_{\odot}~,~~ {\rm and}~~ 0\leq \Omega \leq \Omega_{\rm max}~.
\end{eqnarray}}

\tc{In Fig. \ref{Lhbcontours}, we have shown the $\tilde{L}_{HB}$ contours within the above mentioned region. The blue curve represents $M_{\rm max}(\Omega)$, while the red one corresponds to $M_{\rm mmhb}(\Omega)$.}
\begin{figure}[h!]
\centering
\hspace{0.3cm}
\begin{minipage}[t]{0.47\linewidth}
\begin{center}
\centerline{\includegraphics[scale=0.55]{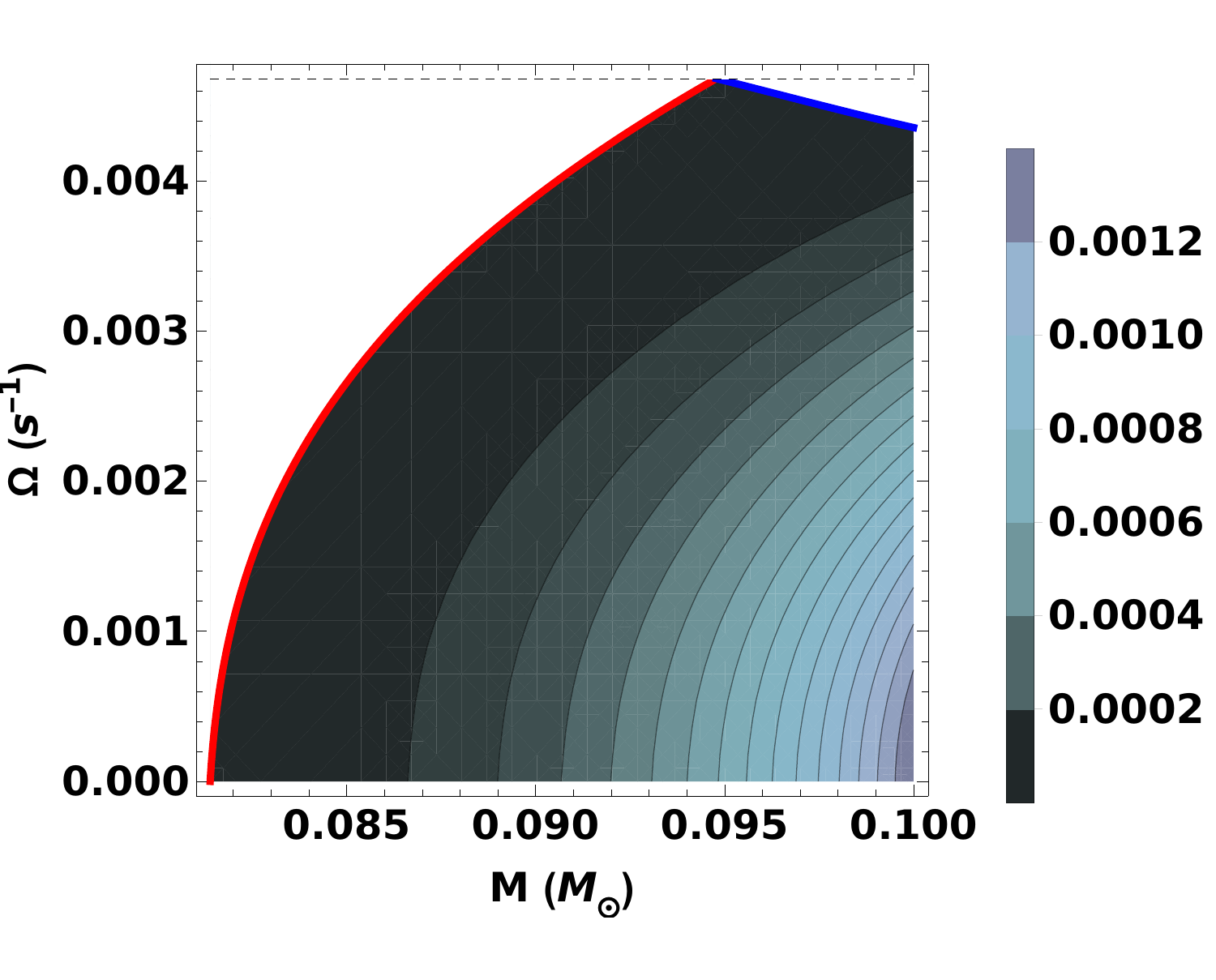}}
\caption{\tc{$\tilde{L}_{HB}$ Contours. Any particular contour of $\tilde{L}_{HB}$ specifies all admissible values of mass $M$ and rotation $\Omega$ of \tc{VLM object}, that acquire the same hydrogen burning luminosity when they reach the main sequence. The validity of this contour plot lies in the admissible range of $M_{\rm mmhb}(\Omega)\leq M\leq M_{\rm max}(\Omega)$ with maximal M and $\Omega$ values of $0.1M_{\odot}$ and $0.0047s^{-1}$. The red curve represents the $M_{\rm mmhb}(\Omega)$, while the blue curve represents the $M_{\rm max}(\Omega)$.}}
\label{Lhbcontours}
\end{center}
\end{minipage}%
\hspace{0.3cm}
\begin{minipage}[t]{0.47\linewidth}
\begin{center}
\centerline{\includegraphics[scale=0.35]{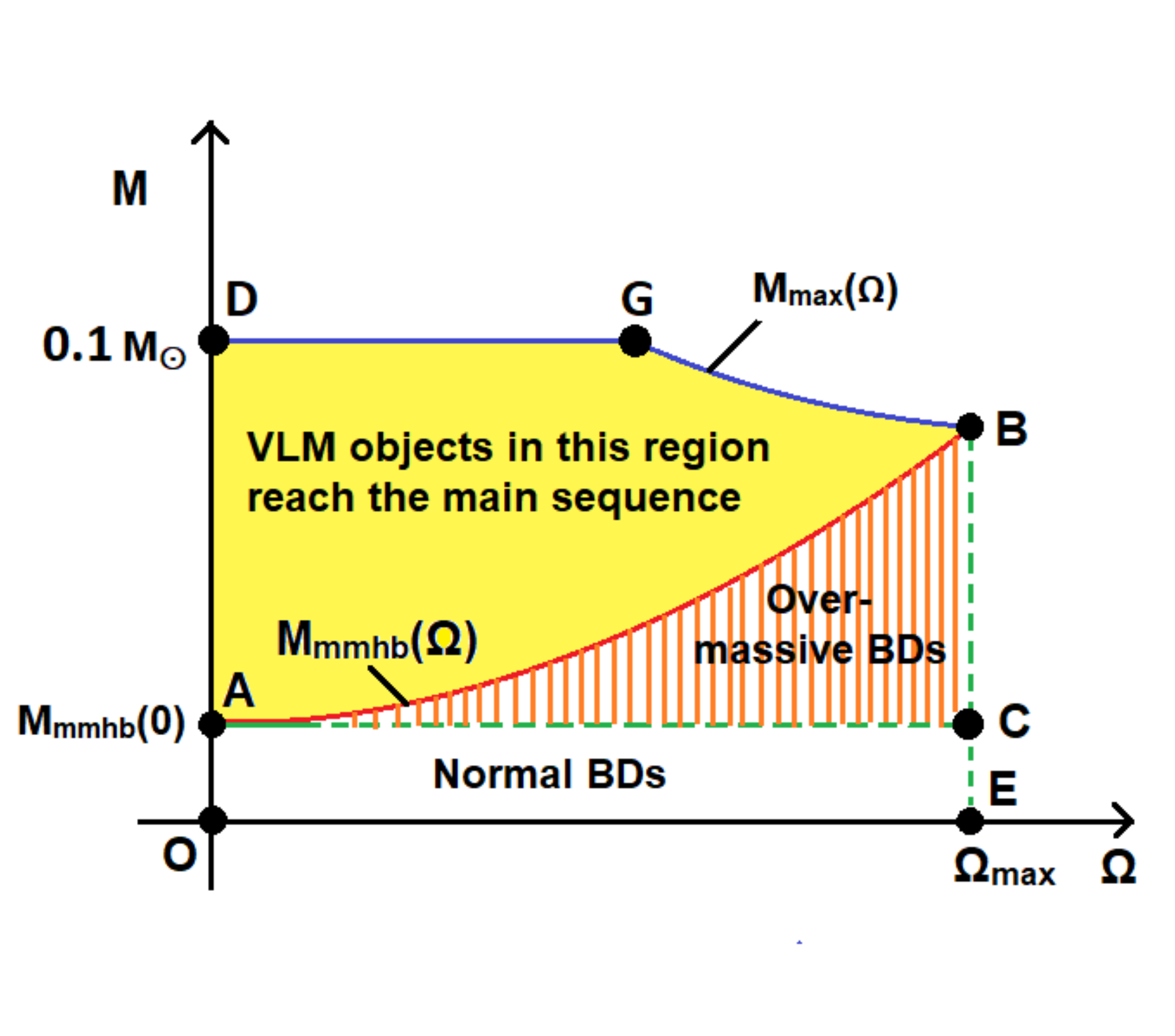}}
\caption{\tc{Schematic diagram of the results obtained in this paper. $M_{\rm mmhb}(0)$ corresponds to $M_{\rm mmhb}(\Omega=0)$. See discussion in text.} 
}
\label{SchematicDiag}
\end{center}
\end{minipage}
\end{figure}

\section{Discussions}
\label{Discussions}

In this paper, we have used a \tc{simplified analytical model} to study the effects of rapid rotation on the 
\mmhb. Our model is inspired by the one due to \cite{BL} to which it reduces to, in the limiting non-rotating case.
Following \cite{Auddy}, we have chosen the values of some of the model parameters from Model D of \cite{Chabrier1992} in order to obtain 
reasonable estimates for the physical parameters. \tc{There are four main results that we have obtained. 
\begin{itemize}
\item
We have found an analytical formula of the \mmhb as a function of the angular speed $\Omega$.  
\item
For a given $\Omega$, we have obtained the mass range \tc{\mmhb($\Omega$) $\leq M \leq$ \mmax($\Omega$)} for \tc{VLM objects} to
evolve into main sequence stars.
\item
We have obtained an upper bound $\Omega_{\rm max} = 0.0047 s^{-1}$ beyond which \tc{a VLM object will not evolve into a MSS}. 
\item
As a by-product of our analysis, we obtained the luminosity of \tc{a VLM object} at the point 
where it reaches the main sequence, as a function of $M$ and $\Omega$. 
\end{itemize}
}
A schematic diagram of the main results in the paper is given
in Fig. \ref{SchematicDiag}. 
Here, the red curve AB represents \mmhb($\Omega$). \tc{The blue curve DG represents the maximal mass of $0.1M_{\odot}$. The blue curve GB represents portion of the \mmax($\Omega$) up to a maximal mass of $0.1M_{\odot}$. The two curves AB and GB intersect at point B.} \tc{Point A
	corresponds to \mmhb value for the non rotating case, labeled as $M_{\rm mmhb}(0)$ in the figure}. The horizontal line AC denotes a constant mass
curve corresponding to the \mmhb for the non-rotating case. Also, \tc{O} corresponds to
origin, while E denotes $\Omega_{\rm max}$. Over-massive BDs (in the region ABC),
have been shown to exist purely due to uniform stellar rotation. \tc{In absence of rotation, BDs lie in the region ACEO, labeled 
as ``Normal BDs.'' The VLM objects in the region DGBA can evolve into MSS.}

Here, we have used a toy model, with a number of assumptions. First, all the thermodynamic relations have been assumed
to remain unaltered in the presence of rotation. This can be justified, as the rotational kinetic energy
$I\Omega^2/2$ with $I$ being the moment of inertia of the deformed \tc{object} computed numerically, can always
be shown to be two orders of magnitude lower than the gravitational potential energy. 
Second, we have considered the effect of uniform stellar rotation
on \tc{VLM object's} evolution. A more realistic situation with differential and time-varying rotation is left for a future study. \te{Third, in our analysis of stellar evolution under constant uniform rotation, the non-conservation of angular momentum has been inherently assumed.}
Finally, our model is polytropic, and does not take account of atmospheric corrections and related details.

However, this simplistic toy model has successfully been able to decode the underlying physics of a rapidly
rotating \tc{VLM object} and has revealed several important limits.

\section{Acknowledgments}
We thank our anonymous referee for various important comments which helped to improve an initial version of the manuscript. 
We acknowledge the High Performance Computing (HPC) facility at IIT Kanpur, India, where the numerical
analysis was carried out.


\begin{thebibliography}{999}
\bibitem[Allard, Homeier, and Fryetag (2012)] {Allard} Allard, F., Homeier, D., and Freytag, B., Phil. Trans. R. Soc. A (2012),
{\bf 370}, 2765.
\bibitem[Auddy, Basu and Valluri (2016)]{Auddy} Auddy, S., Basu, S., and Valluri, S. R., 2016, Adv. Astron., 2016, 5743272.
\bibitem[Banerjee et. al. (2021)]{tapo3} Banerjee, P., Garain, D., Paul, S., et. al., 2021, Astrophys. J. {\bf 910}, 23.
\bibitem[Basri (2000)]{Basri} Basri, G., 2000, Ann. Rev. Astron. Astrophys. {\bf 38}, 485.
\bibitem[Benito and Wojnar (2021)]{BenitoWojnar} Benito, M., and Wojnar, A., Phys. Rev. D \textbf{103}, no.6, 064032. 
\bibitem[Burrows, Hubbard and Lunine (1989)]{Evol2} Burrows, A., Hubbard, W. B., and Lunine, J, I., 1989,
Astrophys. J. {\bf 345}, 939.
\bibitem[Burrows et. al. (2001)]{BL2} Burrows, A., Hubbard, W. B., Lunine, J. I., and Liebert, J., 2001,
Rev. Mod. Phys. {\bf 73}, 719.
\bibitem[Burrows and Liebert (1993)]{BL} Burrows, A., and Liebert, J., 1993, Rev. Mod. Phys {\bf 65}, 301.
\bibitem[Burrows et. al. (1997)]{Burrows2} Burrows, A., Marley, M., and Hubbard, W. B., Astrophys. J. {\bf 491}, 856.
\tc{\bibitem[Chabrier et. al. (1992)]{Chabrier1992} Chabrier, G., Saumon, D., Hubbard, W., B., Lunine, J., I., 1992, Astrophys. J. 391, 817.}
\bibitem[Chabrier and Baraffe (1997)]{CB2} Chabrier, G., and Baraffe, I., 1997, Astron. Astrophys. 327, 1039.
\bibitem[Chabrier and Baraffe (2000)]{CB} Chabrier, G., and Baraffe, I., 2000, Ann. Rev. Astron. Astrophys. {\bf 38}, 337.
\bibitem[Chabrier et. al. (2014)]{Chabrier} Chabrier, G., Johansen, A., Janson, M., Rafikov, R., 2014, in 
Beuther, Henrik, et al. {\tt Protostars and Planets VI}, University of Arizona Press, 619.
\bibitem[Chandrasekhar (1933)] {Chandra2} Chandrasekhar, S., 1933, MNRAS {\bf 93}, 390.
\bibitem[Chandrasekhar (1939)]{Chandra} Chandrasekhar, S., {\tt An Introdution to the Study of Stellar Structure}, 1939,
Univ. of Chicago Press, U.S.A. 
\bibitem[Clarke et. al. (2008)]{Clarke} Clarke, F. J., Hodgkin, S. T., Oppenheimer, B. R., et. al., 2008, MNRAS {\bf 386}, 2009.
\bibitem[D'Antona and Mazzitelli (1985)]{Evol1} D'Antona, F., and Mazzitelli, I., 1985, Astrophys. J. {\bf 296}, 502.
\bibitem[D'Antona and Mazzitelli (1997)]{Evol3} D'Antona, F., and Mazzitelli, I., 1997, 
Memorie della Societa Astronomica Italiana {\bf 68}, 807.
\bibitem[Forbes and Loeb (2019)]{ForbesLoeb} Forbes, J. C., and Loeb, A., 2019, Astrophys. J. {\bf 871}, 2.
\bibitem[Hayashi and Nakano (1963)]{Hayashi} Hayashi C., and Nakano T., 1963, Prog. Theor. Phys. {\bf 30}, 460.
\bibitem[Hurley and Roberts (1964)]{Roberts3} Hurley, M., and Roberts, P. H., 1964, Astrophys. J., {\bf 140}, 583.
\bibitem[Ishii, Shibata and Mino (2005)]{Ishii} Ishii M., Shibata M. and  Mino Y., 2005, Phys. Rev. D, {\bf 71}, 044017.
\bibitem[James (1964)] {James} James, R., 1964, Astrophys. J., {\bf 140}, 552.
\bibitem[Jeans (1928)] {Jeans} Jeans, J. H., 1928, {\tt Astronomy and Cosmogony}, Cambridge University Press, U.K. 
\bibitem[Joergens (2014)]{Joergens} Joergens, V., Ed., 2014, {\tt 50 Years of Brown Dwarfs, From Prediction to 
Discovery to Forefront Research}, Springer, Heidelberg.
\bibitem[Kippenhan (1970)]{KippenRot} Kippenhahn, R., 1970, Astron. Astrophys. {\bf 8}, 50.
\bibitem[Kippenhahn and Thomas (1970)]{KThomas} Kippenhahn, R., and Thomas, H. C., 1970, in Slettebak, A., Ed., {\tt 
Stellar Rotation} Proc. IAU Colloquium 1969, D. Reidel Publishing Company, Dordrecht-Holland.
\bibitem[Kumar (1963)]{SSK} Kumar, S. S., 1963, Astrophys. J. {\bf 137}, 1121.
\bibitem[Maeder (2009)]{Maeder} Maeder, A., 2009, {\tt Physics, Formation and Evolution of Rotating Stars}, Springer-Verlag, Berlin,
Heidelberg.
\bibitem[Marley and Robinson (2015)] {Marley} Marley, M. S., and Robinson, T. D., Ann. Rev. Astron. Astrophys. (2015), {\bf 53}, 279.
\bibitem[Metchev, Heinze and Apai (2015)]{Metchev} Metchev, S. A., Heinze, A., Apai, D., et al. 2015, Astrophys. J. {\bf 799}, 154.
\bibitem[Nakajima et. al. (1995)]{Nakajima} Nakajima, T., Oppenheimer, B., Kulkarni, S. et al., 1993, Nature {\bf 378}, 463.
\bibitem[Nelson, Rappaport and Joss (1986)]{Poly1} Nelson, L. A., Rappaport, S. A., and Joss, P. C., 1986, Astrophys. J. {\bf 311}, 226.
\bibitem[Rappaport and Joss (1984)]{Poly2} Rappaport, S., and Joss, P. C., 1984, Astrophys. J. {\bf 283}, 232.
\bibitem[Rebolo and Zapatero-Osorio (2000)]{Rebolo2} Rebolo, R., and Zapatero-Osorio, M., R., 2000, {\tt Very Low-Mass Stars and 
Brown Dwards}, Cambridge University Press, Cambridge, United Kingdom.
\bibitem[Rebolo, Zapatero-Osorio and Martin (1995)]{Rebolo} Rebolo, R., Zapatero-Osorio, M., R., and Martin, E., 1995, Nature {\bf 377}, 129.
\bibitem[Roberts (1963a)] {Roberts1} Roberts, P. H., 1963, Astrophys. J., {\bf 137}, 1129.
\bibitem[Roberts (1963b)] {Roberts2} Roberts, P. H., 1963, Astrophys. J., {\bf 138}, 809.
\bibitem[Route and Wolszczan (2016)]{Route} Route, M., and Wolszczan, A., 2016, Astrophys. J. Letters {\bf 821} L21.
\bibitem[Salpeter (1992)]{SalpeterAccretion} Salpeter, E. E., 1992, Astrophys. J. {\bf 393}, 258.
\bibitem[Stoeckly (1965)]{Stoeckly} Stoeckly, R., 1965, Astrophys. J., {\bf 165}, 208.
\bibitem[Tannock et. al. (2021)]{Tannock} Tannock, M. E., Metchev, S., Heinze, A., et. al. (2021) Astron. J. {\bf 161}, 224.
\bibitem[Tassoul (2000)]{Tassoul} Tassoul, J-L., 2000, {\tt Stellar Rotation}, Cambridge University Press, Cambridge, United Kingdom. 
\bibitem[Williams, Gizis and Berger (2017)]{Williams} Williams, P. K. G., Gizis, J. E., and Berger, E., 
Astrophys. J. 2017, {\bf 834}, 117.

\end{thebibliography}
\end{document}